\documentclass[seceq,supplement]{ptptex}
\usepackage{graphicx}

\title{Thermodynamics of strong coupling 2-color QCD}
\author{Yusuke~\textsc{Nishida}}
\inst{Department of Physics, University of Tokyo, 
      Tokyo 113-0033, Japan}

\abst{2-color QCD at finite temperature $T$ and chemical potential $\mu$ 
 is revisited in the strong coupling limit on the lattice. 
 The phase structure in the space of $T$, $\mu$ and the
 quark mass $m$ is elucidated with the use of the mean field
 approximation and the large dimensional expansion. 
 We put special emphasis on the interplay among the chiral condensate,
 the diquark condensate and the quark density in the $T$-$\mu$-$m$
 space. Qualitative comparison is made between our results and those
 of recent Monte-Carlo simulations in 2-color QCD.}

\begin{document}
\maketitle

\section{Introduction}
Physics of matter under high baryon density is one of the most
challenging problems in quantum chromodynamics (QCD) both from
technical and physical point of view. Unfortunately, analysing the
finite baryon chemical potential region of 3-color QCD from the first
principle lattice simulations is retarded due to the complex fermion
determinant. However, the situation is different for 2-color QCD in
which the fermion determinant can be made real and the Monte-Carlo
simulations are attainable. Therefore, 2-color QCD provides an unique
opportunity to compare various ideas at finite chemical potential with 
the results from lattice simulations.

One of the major differences between 2-color QCD and 3-color QCD is
that the color-singlet baryon is boson (diquark) in the
former. This implies that the ground state of the 2-color system at
finite baryon density in the color-confined phase is an interacting
boson system, i.e.\ a Bose liquid, although the quark Fermi liquid may
be realized at high baryon density in the color-deconfined phase. How 
this Bose liquid changes its character as a function of the
temperature $T$, the quark chemical potential $\mu$ and the quark
mass $m$ is an interesting question by itself and may also give us a 
hint to understand physics of the color superconductivity in 3-color
QCD in which the crossover from the Bose-Einstein condensate of
tightly bound quark pairs to the BCS type condensate of loosely bound
Cooper pairs may take place.\cite{abuki02}

In this article, we revisit the thermodynamics of the strong
coupling limit of 2-color lattice QCD with chiral and diquark
condensates which was originally studied in Ref.~\citen{dagotto86}. 
Our main purpose is to analyse its phase structure and the interplay
among the chiral condensate $\langle\bar{q}q\rangle$, the diquark
condensate $\langle qq\rangle$ and the baryon density $\langle
q^{\dagger}q\rangle$ as functions of $T$, $\mu$ and $m$. This would give
us a physical insight and useful guide for the actively pursued lattice
QCD simulations of the same system in the weak coupling.\cite{kogut01}

\section{Formulation}
We derive an effective free energy for meson and diquark fields starting
from the lattice action with the single component staggered fermion. 
In the strong coupling limit $g\to\infty$, the gluonic part of the
action vanishes because it is inversely proportional to $g^2$. 
Then the action on the lattice is given by only fermionic part;
\begin{align}
 S=m&\sum_x\bar{\chi}(x)\chi(x)
 +\frac{1}{2}\sum_x\sum_{j=1}^d\eta_j(x)
 \left\{\bar{\chi}(x)U_j(x)\chi(x+\hat{j})
 -\bar{\chi}(x+\hat{j})U_j^\dagger(x)\chi(x)\right\}\nonumber\\
 &+\frac{1}{2}\sum_x\eta_0(x)\left\{\bar{\chi}(x)\mathrm{e}^\mu
 U_0(x)\chi(x+\hat{0})-\bar{\chi}(x+\hat{0})U_0^\dagger(x)
 \mathrm{e}^{-\mu}\chi(x)\right\}\,,
 \label{eq:action}
\end{align}
with
\begin{align}
 \eta_0(x)=1\,,\qquad\eta_j(x)=(-1)^{\sum_{i=1}^j x_{i-1}}\,.
\end{align}
$\chi$ stands for the quark field in the fundamental representation
of the color $\mathrm{SU}(2)$ group and $U_\mu$ is the
$\mathrm{SU}(2)$ valued gauge link variable. 
$d$ represents the number of spatial directions. 

This action has a global $\mathrm{U_V(1)\times U_A(1)}$ symmetry at
$\mu\not=0$ in the chiral limit $m=0$. 
$\mathrm{U_V(1)}$ corresponds to the baryon number conservation,
which is spontaneously broken by the diquark condensation. 
On the other hand, the action has a larger symmetry $\mathrm{U}(2)$ for 
$m=\mu=0$ due to the Pauli-G\"{u}rsey's fermion$-$anti-fermion
symmetry. This is the special feature of 2-color QCD.

Here we summarise how to derive the effective free energy for the meson
and diquark system from the original action Eq.~(\ref{eq:action}).
\begin{enumerate}
\item Large dimensional ($1/d$) expansion is employed in the
 spatial directions in order to facilitate the integration
 over the spatial link variable $U_j$. 
\item Bosonization is performed by introducing the auxiliary
 fields $\sigma$ for $\bar{\chi}\chi$ and $\Delta$ for $\chi\chi$.
 Then the mean field approximation is adopted for the auxiliary
 fields.
\item Integration with respect to $\chi$, $\bar{\chi}$ and
 $U_0$ are accomplished exactly.
\end{enumerate}

After performing these procedure,\cite{nishida03} \ we can write down
the effective free energy for the chiral condensate ($\sigma$) and the
diquark condensate ($\Delta$) as follows; 
\begin{align}
 F_{\mathrm{eff}}[\sigma,\Delta]=\frac{d}{2}\sigma^2
 +\frac{d}{2}|\Delta|^2-T\log\left\{1+4\cosh\left(E_+/T\right)\cdot
 \cosh\left(E_-/T\right)\right\}\,,
 \label{eq:free_energy}
\end{align}
where $E_\pm$ is the excitation energy of (anti) quasi-quarks,
\begin{align}
 E_\pm =\mathrm{arccosh}\left(\sqrt{(1+M^2)\cosh^2\mu
       +(d/2)^2|\Delta|^2}\pm M\sinh\mu\right)
\label{eq:quasi_energy}
\end{align}
with dynamical quark mass $M=m+\sigma d/2$.
In the chiral limit $m=0$ with zero chemical potential $\mu=0$, this 
is a function only in terms of
$\sigma^2+|\Delta|^2$. As a result, the effective free energy is
invariant under the transformation mixing the chiral condensate with the
diquark condensate. This symmetry comes from the Pauli-G\"{u}rsey symmetry
of the original action at $m=\mu=0$. 
The computational details to derive the effective free energy and its
analytical properties can be found in Ref.~\citen{nishida03}.

\section{Numerical Results and Discussions}

\begin{figure}[tp]
 \begin{center}
  \includegraphics[width=.98\textwidth,clip]{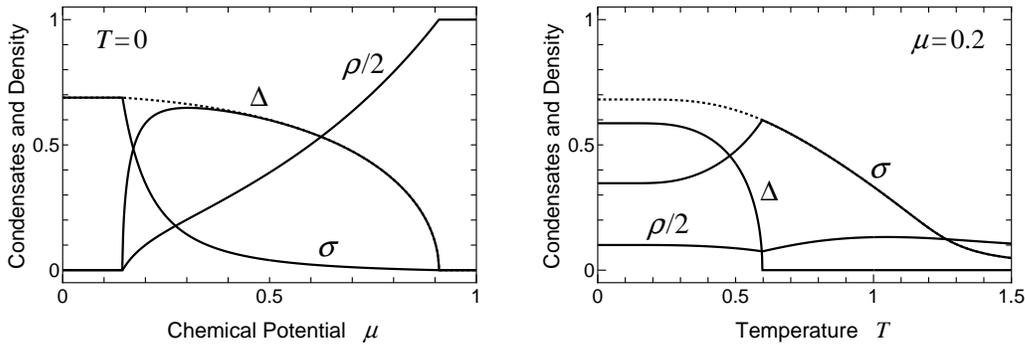}
  \caption{Chiral condensate $\sigma$, diquark condensate $\Delta$ 
  and quark density $\rho$ for $m=0.02$ with $d=3$.
  In the left panel they are plotted as functions of
  the chemical potential $\mu$ at zero temperature.
  In the right panel they are plotted as functions of temperature $T$ 
  for a typical value of $\mu$.
  All the dimensionful quantities are in unit of the lattice
  spacing, which is implicitly understood in other figures. 
  The dotted line indicates a total magnitude of
  the condensates $\sqrt{\sigma^2+\Delta^2}$.}
  \label{fig:massive}
 \end{center}
\end{figure}

We determine the chiral condensate $\sigma$ and the
diquark condensate $\Delta$ numerically by minimising the effective
free energy in Eq.~(\ref{eq:free_energy}). The quark density
$\rho=-\partial F_\mathrm{eff}/\partial\mu$ is also calculated. The
results are shown in Figs.~\ref{fig:massive} for small quark mass
$m=0.02$. 

First we consider the chiral and diquark condensates as functions of 
chemical potential (the left panel of Fig.~\ref{fig:massive}).
There exist two critical chemical potentials, the lower one
$\mu_\mathrm{c}^\mathrm{low}$ and the upper one
$\mu_\mathrm{c}^\mathrm{up}$. 
We can understand the behavior of the condensates as the manifestation
of two different mechanisms: One is a continuous ``rotation'' from the
chiral condensation to the diquark condensation above
$\mu=\mu_{\mathrm{c}}^{\mathrm{low}}$ with the total condensate
$\sqrt{\sigma^2+\Delta^2}$ varying smoothly. 
The other is the ``saturation effect''; quark density forces the 
diquark condensate to decrease and disappear for large $\mu$.

The ``rotation'' can be understood as follows: As we have discussed, 
the effective free energy at vanishing $m$ and $\mu$ has a symmetry 
between the chiral condensate and the diquark condensate. 
The effect of $m$ ($\mu$) is to break this symmetry in the direction of
the chiral (diquark) condensation favored. Therefore at finite $m$, a
relatively large chiral condensate predominantly appears for small $\mu$
region. Once $\mu$ exceeds a threshold value
$\mu_{\mathrm{c}}^{\mathrm{low}}$, the chiral condensate decreases 
while the diquark condensate increases, because the effect of $\mu$
surpasses that of $m$. 

As $\mu$ becomes larger, the diquark condensate
begins to decrease in turn by the effect of the ``saturation''
and eventually disappears when $\mu$ exceeds the upper critical value
$\mu_{\mathrm{c}}^{\mathrm{up}}$ (order of unity for $T=0$).
On the other hand, the quark density $\rho$ increases until the
saturation point ($\rho=2$) where quarks occupy the maximally allowed 
configurations by the Fermi statistics.
Those behaviors of $\Delta$ and $\rho$ are also observed in the recent
Monte-Carlo simulations of 2-color QCD.\cite{kogut01} 

Next we consider the chiral and diquark condensates as functions of
$T$ (the right panel of Fig.~\ref{fig:massive}). At low $T$, both
the chiral and diquark condensates have finite values for $\mu =0.2$. 
The diquark condensate decreases monotonously as $T$
increases and shows a second order transition. On the other hand, the
chiral condensate increases as the diquark condensate decreases so
that the total condensate $\sqrt{\sigma^2+\Delta^2}$ is a smoothly
varying function of $T$. The understanding based on the approximate 
symmetry between chiral and diquark condensates is thus valid. An
interesting observation is that the chiral condensate has a cusp shape
associated with the phase transition of the diquark condensate.

\begin{figure}[tp]
 \parbox[b]{.48\textwidth}
 {\begin{center}
   \includegraphics[width=.475\textwidth,clip]{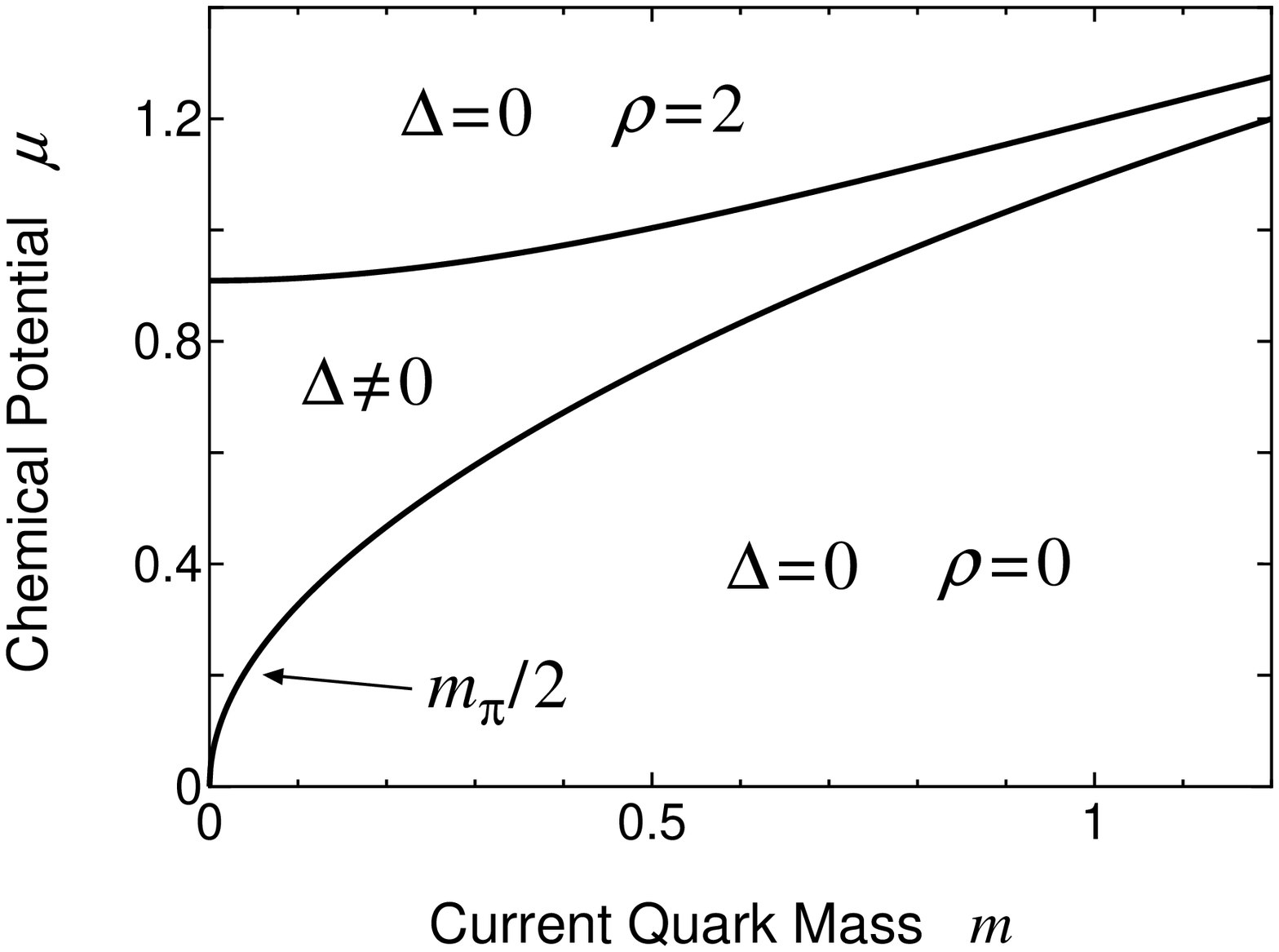}
   \caption{Phase diagram of strong coupling 2-color QCD in the
   $\mu$-$m$ plane at $T=0$. 
   \label{fig:phase_diagram-MuM}}
  \end{center}}
 \hfill
 \parbox[b]{.48\textwidth}
 {\begin{center}
   \includegraphics[width=.48\textwidth,clip]{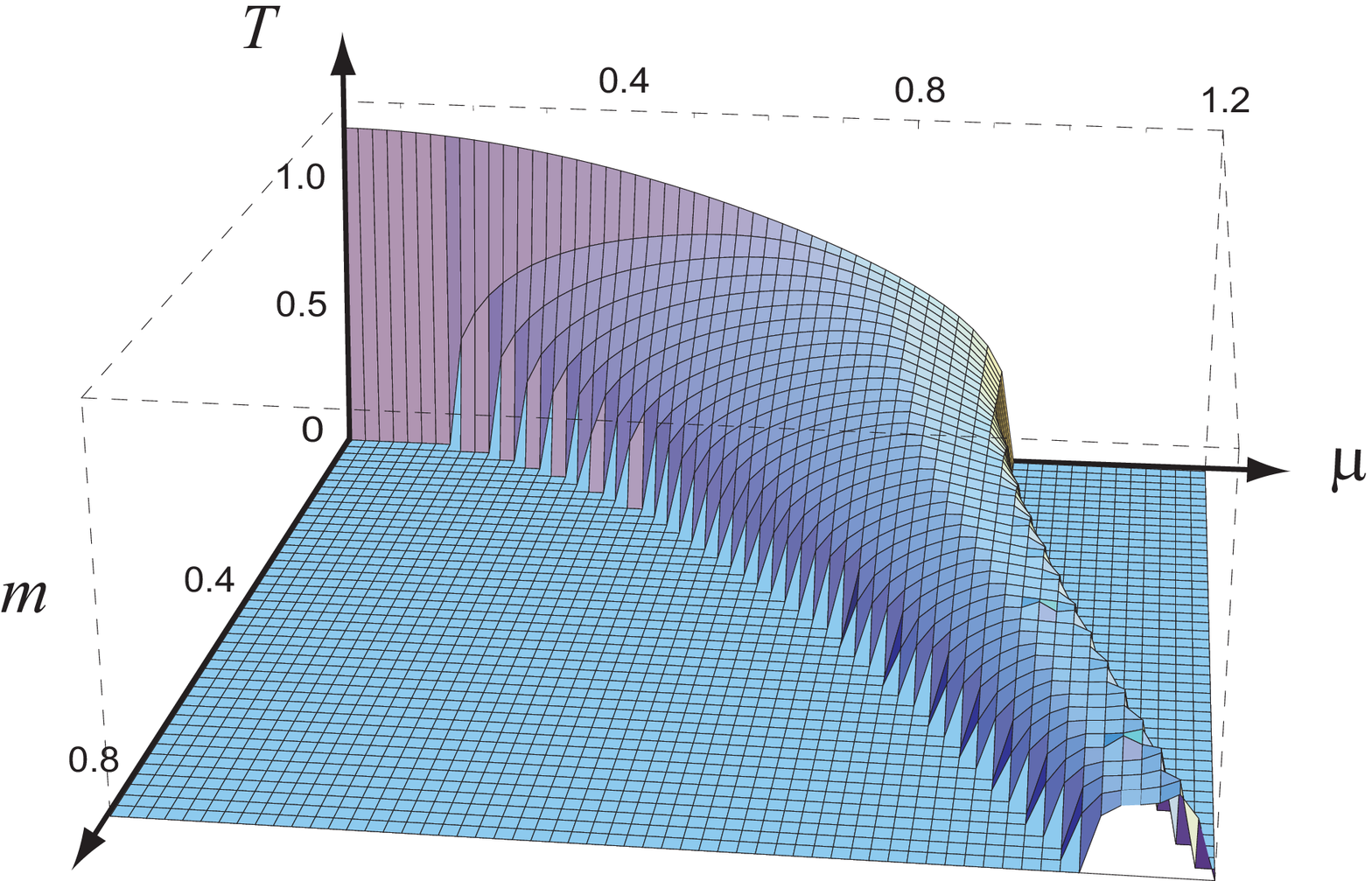}
   \caption{Phase structure of strong coupling 2-color QCD in the
   $T$-$\mu$-$m$ space. The surface separates the region where
   $\Delta=0$ (outside) from the region where $\Delta\not=0$ (inside).
   \label{fig:phase_diagram-3D}}
  \end{center}}
\end{figure}
 
Now we show the phase diagram of the strong coupling 2-color QCD 
in the $\mu$-$m$ plane at $T=0$ in Fig.~\ref{fig:phase_diagram-MuM}.
The lower right of the figure corresponds to the vacuum with
no baryon number present, $\rho=0$. On the other hand, the upper left
of the figure corresponds to the saturated system, $\rho=2$, in which
every lattice site is occupied by two quarks. There is a diquark 
superfluid phase with $\Delta\not=0$ and $0<\rho<2$ bounded by the above
two limiting cases.
It is worth mentioning here that we can see the corresponding system
in the context of condensed matter physics; the hardcore boson Hubbard
model has a similar phase diagram in which superfluid phase is
sandwiched by Mott-insulating phases with zero or full
density.\cite{schmid01} 

Finally, the phase structure
in the three dimensional $T$-$\mu$-$m$ space is shown in
Fig.~\ref{fig:phase_diagram-3D}. The diquark condensate has a
none-vanishing value inside the critical surface and the phase
transition is of second order everywhere on this critical surface.

\section*{Acknowledgements}
This article is based on a work Ref.~\citen{nishida03} in collaboration
with K.~Fukushima and T.~Hatsuda to whom the author is grateful.

\end{document}